\documentclass[english]{iopart}

\usepackage{graphicx}
\usepackage{amssymb}

\newcounter{fig}
\begin{document}

\title[The seven-dimensional fcc lattice]{\Large Lattice Green Functions: the seven-dimensional face-centred cubic lattice}

\author{N. Zenine$^\S$, S. Hassani$^\S$, J.M. Maillard$^\pounds$     
}
\address{\S  Centre de Recherche Nucl\'eaire d'Alger, 
2 Bd. Frantz Fanon, BP 399, 16000 Alger, Algeria}
\address{$^\pounds$ LPTMC, UMR 7600 CNRS, 
Universit\'e de Paris 6, Sorbonne Universit\'es, Tour 23,  
5\`eme \'etage, case 121, 
 4 Place Jussieu, 75252 Paris Cedex 05, France}

\begin{abstract}

We present a recursive method to generate the expansion of the 
lattice Green function of the $d$-dimensional face-centred cubic (fcc) lattice.
We produce a long series for $d =7$.
Then we show (and recall) that, in order to obtain the linear differential equation annihilating
such a long power series, the most economic way amounts to producing the non-minimal order differential
equations. We use the method to obtain the minimal order linear differential equation of the
lattice Green function of the seven-dimensional face-centred cubic (fcc) lattice.
We give some properties of this irreducible order-eleven differential equation. 
We show that the differential Galois group of the corresponding operator is included
in $\, SO(11, \, \mathbb{C})$. This order-eleven operator is non-trivially 
homomorphic to its adjoint, and we give a ``decomposition'' of this order-eleven
 operator in terms of four order-one self-adjoint operators and one order-seven 
self-adjoint operator.
Furthermore, using the Landau conditions on the integral, we forward the regular
singularities of the differential equation of the $\, d$-dimensional lattice and show
that they are all rational numbers.
We evaluate the return probability in random walks in the seven-dimensional fcc lattice.
We show that the return probability in the $\, d$-dimensional fcc lattice decreases
as $\,d^{-2}$ as the dimension $\, d$ goes to infinity.

\vskip .5cm




\end{abstract} 

\section{Introduction}

Many remarkable $\, n$-fold integrals arise in physics. The lattice Green functions are such integrals
and they occur in many problems.
The number of the integration variables is the dimension and the lattice may be simple,
body-centred, face-centred, etc. 
For a review, see~\cite{guttmann-2010} and the references therein.
We focus here on the {\em face-centred cubic} (fcc) lattices.
The corresponding three-dimensional and higher-dimensional
lattice Green functions  have been analyzed
in~\cite{guttmann-2010,joyce-1998,guttmann-2009,broadhurst-2009,koutschan-2013}.

The $\, d$-dimensional lattice Green function of the face-centred cubic (fcc) lattice
reads
\begin{eqnarray}
\label{LGFd}
\hspace{-0.95in}&& \quad \qquad   \qquad     
LGF_d(x) \,\, =\,\,\, \, 
{\frac{1}{\pi^d}}\, \int_0^\pi \cdots \,
 \int_0^\pi \, {\frac{dk_1\,  \cdots \,\,  dk_d}{1\,-x \, \lambda_d}}, 
\end{eqnarray}
where $\lambda_d$ is the structure function:
\begin{eqnarray}
\hspace{-0.95in}&& \quad \qquad   \qquad     
\lambda_d \,\, =\,\,\, { d \choose 2}^{-1} \cdot \, \sum_{i=1}^d \, \sum_{j=i+1}^d 
\cos(k_i) \, \cos(k_j).
\end{eqnarray}

\vskip 0.1cm

For $\, d=\,2$, the integral is the complete elliptic integral of the first kind $\, K$
\begin{eqnarray}
\hspace{-0.95in}&& \quad \qquad   \qquad   
LGF_2(x)\, \,=\,\,\, _2F_1([1/2,\, 1/2],\, \,[1],\, x^2) \, \,= \,\,\, K(x^2).
\end{eqnarray}

For $\, d=\, 3$, Joyce~\cite{joyce-94} has shown that $\,LGF_3(x)$ can be written as
a square of the elliptic integral $K$ with a pullback
\begin{eqnarray}
\hspace{-0.95in}&& \quad \qquad   \qquad   
LGF_3(x) \, \,\,=\, \,\,\, A(x) \cdot \, (K(k_0^2))^2,
\end{eqnarray}
with:
\begin{eqnarray}
\hspace{-0.95in}&& \qquad   \qquad     
A(x) \, \,= \, \,\, \,3^{3/2} \cdot \, (3+x)^{-3/2}  \cdot \, (2 - (1-x)^{1/2}),
 \\
\hspace{-0.95in} &&  \qquad \qquad       
  k_0^2 \,\,=\,\, \,\,{1 \over 2}  \, \,\, 
-{\sqrt{3} \over 2} \cdot \, (3+x)^{-3/2} \cdot \,  (4 x \, + (3-x)\,(1-x)^{1/2}).
\end{eqnarray}

\vskip 0.1cm

For higher values of $\, d$, no closed form expression for (\ref{LGFd}) is known.
The integral being $\,D$-finite~\cite{Stanley,lipshitz-89}, 
it is possible to study some of its properties
via its linear differential equation (ODE).

\vskip 0.1cm

The linear ordinary differential equation annihilating the integral (\ref{LGFd}),
for $\,d= \, 4$, was obtained by Guttmann~\cite{guttmann-2010}.
By a change of variables and integration, he succeeded 
to eliminate two variables, and was left with a two-dimensional integral which was
expanded as a power series in the variable $\, x$, and integrated term-by-term. 
The linear ODE (the corresponding operator is denoted $\, G_{4}^{4Dfcc}$) 
is of order {\em four} and is a {\em Calabi-Yau equation} (number 366 
in the list of Almkvist et al.~\cite{TablesCalabi}).

To obtain the linear ODE for $\,d=\,5$, Broadhurst~\cite{broadhurst-2009} expanded the
integrand in (\ref{LGFd}) as a power series, then used the multinomial theorem
to re-expand the powers of the structure function. The integration of each term
is straightforward, but the computation demands nine summations.
The linear ODE (associated with operator $ \, G_{6}^{5Dfcc}$) is of order {\em six}, 
but we do not have the presence of maximum unipotent 
monodromy (MUM)\footnote[1]{Maximum unipotent 
monodromy~\cite{guttmann-2010,TablesCalabi, 2011-calabi-yau-ising, Lian,Hayama}
 means that the local exponents at ,e.g., the origin are all equal.}.

Koutschan~\cite{koutschan-2013} used the method of {\em creative 
telescoping}~\cite{zeilberger-1990,chyzak-2000,koutschan-2010},
 and was able to find the ODE for $\,d= \, 6$.
This linear ODE (associated with operator $ \, G_{8}^{6Dfcc}$) 
is of order {\em eight}, and lacks MUM.
Koutschan seemed pessimistic in~\cite{koutschan-2013} on the possibility 
to obtain the linear ODE for $\,d= \, 7$ 
with the current methods (of creative telescoping) and computational ressources.

\vskip 0.1cm

In the sequel, we give the linear differential equation for the 
{\em seven-dimensional lattice Green function} of the {\em face-centred cubic} lattice
with some of its properties.
We focus on some properties (singularities, order, differential Galois groups, ...) 
shared by the linear ODE of the {\em $d$-dimensional lattice Green function} of the 
(fcc) lattice, $d \le 7$.

\vskip 0.1cm

The paper is organized as follows. 
Section \ref{generationseries} contains the method of generation 
of the series corresponding to (\ref{LGFd}) for any $\, d$.
This amounts to using a {\em recursive relation} which is less ``summation consuming'' than
the {\em direct expansion using the multinomial theorem}.
For $\,d= \, 7$ a long series is obtained. 
Section \ref{recalls} recalls some tools on how, and why, a linear ODE 
can be obtained with series of {\em fewer terms} than the one necessary to find 
the minimal order ODE. 
These tools have, in previous works, allowed us to obtain many linear ODEs
of quite high orders, requiring very long series of some multifold integrals 
of the magnetic susceptibility of the 
Ising model~\cite{2004-chi3,2005-chi4,2009-chi5,2010-chi6},
or some integrals of the ``Ising-class''~\cite{2007-PhiD-integrals,2007-PhiH-integrals}.
Section \ref{theseven} presents the {\em order-eleven} linear differential equation 
annihilating (\ref{LGFd}) for $\, d= \, 7$ with some of its properties. 
Section \ref{decomp} deals with some analysis previously~\cite{2014-DiffAlg-LGFCY}
 carried out on the 
linear differential operators $\,G_{6}^{5Dfcc}$ and $ \, G_{8}^{6Dfcc}$,
for the five and six-dimensional fcc lattices,
showing that the differential Galois groups of these 
operators are, respectively, included in the 
$\, SO(6, \, \mathbb{C})$ and $\, Sp(8, \, \mathbb{C})$ groups. As 
a consequence these operators have a ``canonical decomposition''~\cite{2014-Decomp-Special}. 
The order-eleven operator for the seven-dimensional fcc lattice Green function
is seen to be {\em non-trivially homomorphic to its adjoint}, and we give 
a {\em decomposition of this order-eleven
 operator in terms of four order-one self-adjoint operators and one order-seven 
self-adjoint operator}.
The singularities of the multifold integral (\ref{LGFd}), for arbitrary $\,d$, 
can be obtained with ``Landau conditions method'', and are 
the subject of section \ref{landau}.
Some comments and speculations are given in section \ref{specul},
and we finally sum-up our results in section \ref{summary}.

\vskip 0.1cm

\section{The seven-dimensional fcc lattice: series generation}
\label{generationseries}

A straightforward way to get $\,LGF_d(x)$ would be to expand the integrand in the variable
$x \cdot \lambda_d$
\begin{eqnarray}
\hspace{-0.95in}&& \qquad \quad \quad \quad \quad  \quad 
 {\frac{1}{1\, -x  \cdot \, \lambda_d}} \, \,=\,\,\, \, \sum_{n=0} \, x^n \cdot \, \lambda_d^n, 
\end{eqnarray}
and to use the multinomial theorem to re-expand $\lambda_d^n$, leading to 
$(d+1)(d-2)/2$ summations.

\vskip 0.1cm

We present, in the sequel, how we have generated our series.
Let us define the variable $\,z =\, {1 \over 4} \,x\, { d \choose 2}^{-1} $, and 
the symbol $< \cdot >$ to mean that the integration 
on the variables $\,k_j$, occurring in the integrand, has been performed
(with the normalization $\pi^d$). 

We define
\begin{eqnarray}
\hspace{-0.95in}&& \qquad   \quad    \, \,
\zeta_d \,\, = \, \,\,  \,  
\sum_{i=1}^d \, \sum_{j=i+1}^d \, \cos(k_i) \cdot \, \cos(k_j), \, \,\quad   \quad     
\sigma_d \,\, = \, \, \,\,   \sum_{i=1}^d \, \cos(k_i), 
\end{eqnarray}
and introduce
\begin{eqnarray}
\label{Tdfirst}
T_d \left(n, j \right) \, \,  \,  \,=\, \,\, \, \,  
 4^{n+j} \, < \zeta_d^n \cdot \sigma_d^{2 j} >, 
\end{eqnarray}
in terms of which $\, LGF_d(x)$ will be given by
\begin{eqnarray}
\hspace{-0.95in}&& \qquad \quad  
LGF_d(x) \,\,\,   =\, \, \, \,  \sum_{n=0} \, z^n \cdot \, < 4^n \, \zeta_d^n > 
\,\, \, \, = \, \, \,\, \sum_{n=0} \, z^n \cdot \, T_d(n,0). 
\end{eqnarray}
The intermediate variables $\,\zeta_d$ and  $\,\sigma_d$ have the obvious
coupled recursions
\begin{eqnarray}
\label{recurzeta}
\hspace{-0.95in}&& \qquad   \qquad     
\zeta_d \,\, \,=\, \,\,\,\,  \zeta_{d-1} \,\,\, + \, \cos(k_d) \cdot \sigma_{d-1}, \\
\label{recursigma}
\hspace{-0.95in}&& \qquad   \qquad  
\sigma_d \,\,\, = \,\, \, \, \,\sigma_{d-1} \,\, \,+ \, \cos(k_d). 
\end{eqnarray}
The right hand side of both (\ref{recurzeta}), (\ref{recursigma}) are substituted
into (\ref{Tdfirst}) to give
\begin{eqnarray}
\hspace{-0.95in}&& \quad \, \, 
 T_d (n, j)\,\,  \,= \\
\hspace{-0.95in}&& \quad \quad  \quad \, \,  \quad  
  4^{n+j} \, \sum_{m=0}^n \, \sum_{p=0}^{2 j} \,
{ n \choose m} \, { 2 j \choose p } \, < \zeta_{d-1}^{n-m}
 \cdot \, \sigma_{d-1}^{p+m} > \cdot\,  < \cos(k_d)^{2j-p+m} >, 
\nonumber 
\end{eqnarray}
where the integration 
\begin{eqnarray}
\hspace{-0.95in}  
\quad  < \cos(k_d)^{2j-p+m} > \, \, \,  \,= \, \,\,  
 \, (1 \, + (-1)^{p+m}) \cdot {\frac{4^{-j+p/2-m/2}}{2}} \cdot
{ 2j-p+m \choose j-p/2+m/2}, 
\end{eqnarray}
is straightforward, and  where
\begin{eqnarray}
\hspace{-0.95in}  \qquad  \quad 
 < \zeta_{d-1}^{n-m} \cdot \sigma_{d-1}^{p+m} > \, \,\, \,=\, \,\, \, 
 4^{-n-p/2+m/2} \cdot T_{d-1} \, ( n-m, {1 \over 2}(p+m)), 
\end{eqnarray}
is a consequence of the definition.

The recursive relation giving $\, T_d (n, j)$ reads
\begin{eqnarray}
\label{Tdnj}
\hspace{-0.95in} \,\,   T_d (n, j) \, \,=\,\, \, \, 
 \sum_{p=0}^n \, \sum_{q=q_1}^{q_2} \,
{ n \choose p} \, { 2 j \choose 2 q+p-n } \,
 { 2 n\, +2 j\, -2 p\, -2 q \choose n+j-p-q} \cdot \, T_{d-1} (p, q), 
\\
\fl \qquad \quad \quad q_1\,  =\,\,   [(n-p+1)/2], 
\quad \quad \quad \quad  q_2 \, = \,\,  [(n-p+2 j)/2], 
\end{eqnarray}
where $[x]$ is the integer part of $x$.
To start the recursion, one needs: 
\begin{eqnarray}
\label{T2nj}
\hspace{-0.95in} && \qquad \, \, 
 T_2 (n, j)\,)\,  \,=\,\, \, \, \sum_{p=p_1}^{p_2} \,
{ 2 p  \choose p} \, { 2 j \choose 2 p-n } \, { 2 n+2 j-2 p \choose n+j-p}, \\
\hspace{-0.95in} && \qquad  \qquad \quad \quad 
p_1 \, = \,\,  [(n+1)/2], \quad \quad \quad  \quad  \quad   p_2 \, =\, \,  [(n+2 j)/2].
\end{eqnarray}
Note that the summation on $\, p$ in (\ref{T2nj}) can be carried out to obtain a closed 
form expression of $T_2(n, j)$, in terms of binomials, $_4 F_3$ and $_5 F_4$ hypergeometric 
functions with unity as argument.

\vskip 0.1cm

While the use of the multinomial theorem leads to $\,(d+1)(d-2)/2$ summations, there are
{\em only} $\,2 d\, -3$ {\em summations in the recursive relation} (\ref{Tdnj}). 

To obtain, with recursion (\ref{Tdnj}),  the coefficient 
of $\, x^n$ in $\, LGF_d(x)$, one remarks that one needs {\em all the quantities} 
$\, T_{\tilde{d}}(p,q)$ 
of the {\em lower dimensions} $\, \tilde{d} <\, d$, with $\, p,\, q\, \le\, n$.
The recursion (\ref{Tdnj}) is {\em still clearly superior  compared to the direct
expansion}.
For instance, to generate 106 terms necessary to obtain the operator $\, G_{6}^{5Dfcc}$,
several CPU days of calculations are mentioned in~\cite{broadhurst-2009}.
With our recursive method, this requires {\em only} 31 seconds on a desk computer.

\vskip 0.1cm

For $\,d=\, 7$ we have generated 460 terms:
\begin{eqnarray}
\label{seriesLGF7}
\hspace{-0.95in} && 
 \quad \, \,   LGF_7(x) \, \,=\, \, \, \,
1\,\, +{\frac {1}{84}}\,{x}^{2}\, \,+{\frac {5}{1764}}\,{x}^{3}\, \,
+{\frac {263}{197568}}\,{x}^{4}\, +{\frac {1355}{2074464}}\,{x}^{5}\, \, \, \, + \,\,\, \cdots
\end{eqnarray}
To generate 460 terms our recursive method
has required some 12 hours on a desk computer.

\vskip 0.1cm

\section{Recalls: minimal order ODE versus non-minimal order ODE}
\label{recalls}

To find the linear ODE annihilating a series $\, S(x)$, like the expansion 
of (\ref{LGFd}), we write the linear ODE as~\cite{2009-chi5}:
\begin{eqnarray}
\label{LQD}
\hspace{-0.95in}&& \quad \qquad   \qquad   
        L_{QD} \, \, \,  = \,\,\,  \, \, 
 \sum_{i=0}^{Q}\,  \Bigl(\sum_{j=0}^{D}\,  a_{i j} \cdot \,  x^j \Bigr)
 \cdot \, \Bigl(x\, {{\rmd} \over {\rmd x}}\Bigr)^i.
\end{eqnarray}
With the condition $\, a_{Q 0}\, \ne \, 0$ (resp. $\, a_{Q D}\, \ne \, 0$), we 
automatically satisfy the requirement for $\, x=\, 0$ (resp. $\,x=\,\infty$) 
to be a {\em regular} singular point.
Note that since $\, S(x)$ is the expansion of an integral with algebraic integrand,
the ODE will be globally nilpotent and therefore Fuchsian \cite{2009-global-nilpotence}.

The polynomials in front of the (homogeneous) derivative  $x\, {{\rmd} \over {\rmd x}}$ have
all the {\em same} degree $\,D$, and the problem amounts 
to solving 
the linear set of equations $\,L_{QD}\left(S(x)\right)\,=\, \, 0$
in the coefficients $\,a_{i j}$.
This method is called by some authors ``guessing method''.

The linear differential operator $\,L_{QD}$, defined by $\, L_{QD}(S(x))\,=\, \, 0$ 
for a given $\,S(x)$, is {\em not unique} if one does not require $\,L_{QD}$ to be of 
minimal order. The advantage of seeking for the {\em non-minimal} order $\,L_{QD}$ 
is that the number of unknown coefficients
to be found can drop dramatically~\cite{2005-chi4}.
Among all these linear ODEs, one is of minimal order $\,q$  
and it is unique (its corresponding degree will be denoted by $\,D_m$). 
In terms of linear differential operators, 
the minimal order differential operator appears as a right-factor in the 
factorization of the 
non-minimal order linear differential operators. The minimal order 
linear ODE may contain a very large number of {\em apparent singularities}, 
and can, thus, only be determined from a very  large number of series 
coefficients (generally $\,N_m =\, (q+1)(D_m+1)$ terms are needed).
Other (non-minimal order) linear ODEs, because they carry polynomials 
of smaller degrees, may require {\em fewer series coefficients} in order to 
be obtained. 

The order of the minimal order ODE can, in fact, be
obtained, from the non-minimal order ODEs, from a remarkable formula, that we reported 
in~\cite{2008-experimental-mathematics-chi},
for which we have no proof, but which has been found to work 
for all the cases we have considered.
This ``{\em ODE formula}'' reads:
\begin{eqnarray}
\label{ODEform}
\hspace{-0.95in}&& \quad \qquad   \qquad   
N \, = \, \, \,\, m \cdot Q \, \,  + q \cdot D \, \,  -C \,\,
=  \,  \, \,\, (Q+1) \cdot \, (D+1)\, \, \, -f.
\end{eqnarray}
The ``{\em ODE formula}'' (\ref{ODEform}) should be understood as follows:
For a long series $ \, S(x)$ we use three (or more) sets of $ \, (Q, \,  D)$ and
solve $\, L_{QD}\left(S(x)\right)\,=\, \, 0$.
From this we obtain, for each pair $(Q, D)$, the value of the parameter $f$, which is the 
number\footnote[2]{Solving $\,L_{QD}(S(x))\,=\, \, 0$ 
(e.g. by term by term) will fix all the
coefficients but leaves $f$ coefficients unfixed among the $\,(Q+1)(D+1)$ ones.
These are all independent ODE solutions for given $\,Q$ and $\,D$.}
of independent solutions (if $\, f>\, 0$, otherwise we increase 
$\,Q$ and/or $\,D$) for each pair $\,(Q,D)$.
These values ($Q$, $\,D$, $\,f$) are then used to determine 
$\,m$, $\,q$ and $\,C$ in (\ref{ODEform}).
In all cases we have investigated, the parameter $\,q$ is the order of the
minimal order linear ODE that annihilates  $\,S(x)$.
The parameter $m$ is the number of singularities (counted with multiplicity) 
excluding any apparent singularities, and the singular point $\,x=\,0$ 
which is already taken care of  by the use of the homogeneous derivative 
$\,x {\rmd \over \rmd x}$. The parameter $\,C$ is related 
to the degree $\, D_{app}$ of the polynomial, in front of 
the highest derivative of the minimal order ODE, carrying the apparent
singularities. $D_{app}$ reads~\cite{2009-chi5}:
\begin{eqnarray}
\label{Dapp}
\hspace{-0.95in}&& \quad \quad   \qquad    \qquad   
D_{app} \,\, \,=\, \,\, \,(m-1) \cdot \, (q-1)\,\, \,-C\,\,\,  -1.
\end{eqnarray}
This formula is a direct consequence of (\ref{ODEform}) and is obtained
for the values $Q=q$, $D=m+D_{app}$ and $f=1$ which define the minimal order ODE.
The degree of the apparent 
polynomial is {\em known even if} the underlying minimal order 
ODE has not been produced.

For a non-minimal order ODE ($Q > \,q$), there is a degree $\,D$ such that
 $\,N_0 = \,(Q+1)(D+1)$ is less $\,N_m$.
Among all these non-minimal ODEs, there is one requiring 
the {\em minimal number} of terms to be found,
and which we call the ``optimal'' ODE (in a ``computational sense'').
For instance, for $\,\tilde{\chi}^{(5)}$, the ``ODE formula'' 
reads~\cite{2009-chi5,2008-experimental-mathematics-chi}:
\begin{eqnarray}
\hspace{-0.95in}&& \quad \quad   \qquad   
N \,\, =\,\,\, \, \, 72  \,Q \,+ 33 \,D \,\, - 887
 \,\, \,  \,=\,\,\,\,  (Q+1) \cdot \, (D+1)\,\, -f.
\end{eqnarray}
The ``optimal'' ODE, i.e. the linear ODE that requires the {\em minimum number}
 of terms in the series to be found, has the triplet 
$\,(Q_0,\, D_0,\, f_0)\, =\,\, (56,\, 129, \,8)$ which corresponds to the minimum 
number $\,N_0 =\, 7402$. The minimal order ODE corresponds to
 the triplet $\,(33,\, 1456,\, 1)$, 
and requires $\,N_m =\, 49537$ series terms.

If we compare both numbers of series terms (for the optimal ODE and the minimal order ODE)
for the known lattice Green function of the $\,fcc$ lattice, one obtains that
the minimal order ODE (resp. optimal ODE) requires 40 (resp. 40) terms for $\,d=\,4$,
98 (resp. 88) terms for $\,d=\,5$ and 342 (resp. 228) for $\, d=\,6$.
The ODE formula for the $\,d=\,6$ fcc lattice ODE reads
\begin{eqnarray} 
\hspace{-0.95in}&& \qquad \qquad 
N \,\, =\, \, \,\, 12  \,Q \, \,+ 8 \,D \, \,- 51
  \,\, \,\,\,=\,\,\, \, (Q+1) \cdot \, (D+1)\,\, -f, 
\end{eqnarray}
which gives with (\ref{Dapp}) an apparent polynomial of degree 25. The gain 
in the number of series terms required to find
 the ``optimal'' linear ODE, instead of the minimal order
 linear ODE, is 114. 
The drop in the number of terms is a consequence of the disappearance of the apparent
singularities.
We expect $\,D_{app}$, and the gain, to be higher for the linear ODE of the
seven-dimensional fcc lattice. 

Note that, once the optimal ODE has been obtained, we may produce enough terms of the series
to get\footnote[1]{It is obvious that the minimal order ODE is the greatest common right divisor (gcrd) of 
all the non minimal order ODEs. 
However, this can hardly be used for the large order ODEs, especially that, for efficient computations, 
the non minimal order ODEs are obtained modulo primes.}
the minimal order ODE. 
The calculations are performed {\em modulo various primes}, and the minimal order ODE is obtained by 
rational reconstruction~\cite{2009-chi5, 2010-chi5-exact, wang-ratrec, monagan-ratrec, col-enc-ratrecon}.

Having recalled the tools that allow us to produce the linear ODE with the fewer number of terms
in the series, we now look for the linear ODE that annihilates the
generated series (\ref{seriesLGF7}).

\vskip 0.1cm

\section{The seven-dimensional fcc lattice: differential equation}
\label{theseven}

With the generated 460 terms, we seek (modulo a prime) for a linear ODE by solving 
 $ \, L_{QD}\left(S(x)\right)\,=\, \, 0$, steadily increasing $\, Q$ and/or $\, D$ 
(with $\, (Q+1)(D+1) < 460$) until we get a positive answer (the parameter $\, f > \, 0$).
When this happens, we produce four ODEs, i.e. we have four sets $\, (Q, D, f)$,
that we use in (\ref{ODEform}), to obtain the ``ODE formula'':
\begin{eqnarray}
\label{ODEformLGF7}
\hspace{-0.95in} && \qquad \qquad 
N \,\, =\,\, \,\,  15 \,Q\,\, +11 \,D\,\, -94 
\, \,\, \,=\,\, \,\, (Q+1) \cdot \, (D+1)\,\, -f.
\end{eqnarray}
The minimal order ODE (underlying our non-minimal order ODEs) is of {\em order eleven} and 
has fifteen singularities (with multiplicity). The degree of the apparent polynomial is 45. 
With this formula, one knows\footnote[3]{Run through the integers
 $\, Q >\,  q =\,  11$, $\,  D >\,  m\,  =\,  15$
with $\,f >\,0$, in (\ref{ODEformLGF7}) to obtain the minimal value of $\, N$.}
 that the optimal ODE corresponds to $\, Q_0 = \, 16$, 
$\, D_0 =\,  22$, $\, f_0 = \, 3$, and requires 391 terms to be found. We 
have then 69 terms as a check. The parameter $f_0 =\,  3$ means 
that we have three independent linear ODEs with order $\, Q_0 =\,  16$ and
degree $\, D_0 = \, 22$.

\vskip 0.1cm

Even if this ``optimal'' linear ODE is only known {\em modulo primes},
 and is of non minimal order, one may 
{\em recognize the singularities}, and compute the local exponents at any singular point. 
At $\, x=\, 0$, the local exponents are $\, \rho=\, 0$ seven times, 
$\, \rho=\, 1$ three times and $\, \rho=\, 3$. 
The five extra and ``spurious'' solutions have no local exponents over the rationals.
The ``ODE formula'' shows that (with $\, Q =\, 11$ and $\, f= \, 1$) 
the minimal order linear ODE needs 732 terms,
which is quite higher than our generated series of 460 terms.
The ``optimal'' ODE (of order 16) is used to generate the terms 
necessary to obtain the minimal order linear ODE.
This process (i.e. use the series of 460 terms known in exact arithmetic, 
obtain the ``optimal'' linear ODE modulo
a prime, then the minimal order ODE) is repeated for many primes 
until the linear ODE can be reconstructed
in {\em exact arithmetic}\footnote[1]{The polynomial of degree 45, carrying apparent singularities, required 
23 primes, and 33 primes were sufficient to reconstruct all the other polynomials.}. 
We call the corresponding linear differential operator $\, G_{11}^{7Dfcc}$ (with $D_x$ the derivative $d/dx$)
\begin{eqnarray}
\hspace{-0.95in} && \quad  \quad \qquad  \qquad 
G_{11}^{7Dfcc}\, \,=\,\,\,\,\,
 \sum_{k=0}^{11} \,Q_k(x) \cdot \, P_{k}(x) \cdot \, D_x^{k}, 
\end{eqnarray}
with
\begin{eqnarray}
\label{singLGF7}
\fl \qquad Q_{11}(x) \,\,=\,\,\, 
{x}^{8} \cdot \, (x+7)^{4} \, (x-1) \, (x+35)  \, (x+21)  \, (x+14)   \, (3+x) \,  (x-21) 
\nonumber \\
\fl \qquad \qquad \qquad \quad  \quad  \times (5\,x+7)  \, (2\,x+7) 
 \, (2\,x+21)  \, (5\,x+63)  \, (3\,x-7), 
\end{eqnarray}
and
\begin{eqnarray}
\fl \qquad  Q_{10}(x) \, \,=\,\, \, 
 {x}^{7} \, (x+7)^{3}, \quad \quad \, \, Q_{9}(x) \,\, =\,\,  {x}^{6} \, (x+7)^{2}, 
\quad\quad \, \,
Q_{8}(x) \,=\, {x}^{5} \left( x+7 \right), 
\nonumber \\
\fl \qquad  Q_{7}(x) \,=\, {x}^{4}, \quad\, \, \quad \, Q_{6}(x) \,=\, {x}^{3},
 \quad \quad \, \,\, Q_{5}(x) \,=\, {x}^{2},
 \quad \quad \, \, \,Q_{4}(x) \,=\, {x}, 
\nonumber \\
\fl \qquad  Q_j(x)=\, 1, \quad \quad  \quad  \quad \quad \quad \,\, j=\, \, 0, \cdots,\,  3 
\end{eqnarray}
The polynomials $\, P_{11},\,  P_{10}, \cdots, \, P_0$ are respectively, of degree
45, 57, 58, 59, 60, 60, 60, 60, 60, 59, 58, 57. 
The polynomial $\, P_{11}(x)$ is given in \ref{appendixApp}. The other polynomials $\, P_j(x)$
are too large to be given here.

\vskip 0.1cm

The indicial equation, at the singularity $\, x=\, 0$, is 
$\,\, \rho^7 \cdot \, (\rho-1)^3 \cdot \, (\rho-2)\, =\, 0$.
The local exponents, at the singularity $\, x =\, -7$, are $\, \rho=\, 3/2$, $\, \rho=\, 5/2$,
three times $\, \rho=\, 2$ and $\rho = \, 0,\, 1,\, 3, \,\cdots, \,6$. 
At $\, x=\, \infty$, the local exponents are $\, \rho=\, 7/2$ and $\rho = 1, 2, \cdots, 10$.
At all the other regular singularities, the local exponent are $\, \rho=\, 5/2$ and
$\rho = 0, 1, \cdots, 9$.
The roots of the degree-45 polynomial $\, P_{11}$ are {\em apparent} singularities
with local exponents $\,\rho = \,0,\, 1, \,\cdots, \,9,\, 11$.

\vskip 0.1cm

The formal solutions of the order-eleven operator $\, G_{11}^{7Dfcc}$ 
are as follows.
There is one solution beginning as $\, x^2 \, + \cdots$, 
\begin{eqnarray}
\label{sollog0}
\hspace{-0.95in} \qquad \quad 
S_2\,  \,=\,\, \,   \,  \,   
{x}^{2}\, \,  \,    -{\frac {493}{2970}}\,{x}^{3}\, \,   \, 
  +{\frac {360763}{2910600}}\,{x}^{4}\,
  \,  \,   -{\frac {146728781}{13335840000}}\,{x}^{5} \,    \, \,  +  \,   \,  \cdots
\end{eqnarray}
Three solutions come as
\begin{eqnarray}
\label{sollog2}
\fl \qquad S_1\, \ln(x)^2 \,  \,  + [x^2]\, \ln(x) \, \,   + [x^3], 
\nonumber \\
\fl \qquad S_1\, \ln(x)\,  \,  + [x^3],  \\
\fl \qquad  S_1 \,   \,=\, \, \,    
x\, \, \,    -{\frac {55}{1008}}\,{x}^{2}\, \,  +{\frac {10097}{190512}}\,{x}^{3}\,  \, 
+{\frac {296669}{85349376}}\,{x}^{4}\, \,  +{\frac {514600171}{67212633600}}\,{x}^{5}
 \, \,  + \, \,   \cdots,  
\nonumber 
\end{eqnarray}
and seven solutions as
\begin{eqnarray}
\label{sollog6}
\fl \quad S_0\, \ln(x)^6 \, + [x]\, \ln(x)^5 \, + [x]\, \ln(x)^4 \, + [x]\, \ln(x)^3 \, 
+ [x^2]\, \ln(x)^2 \, + [x^2]\, \ln(x) \, +  [x^3], 
\nonumber \\
\fl \quad S_0 \, \ln(x)^5 \, + [x]\, \ln(x)^4 \, + [x]\, \ln(x)^3 \, 
+ [x^2]\, \ln(x)^2 \, + [x^2]\, \ln(x) \, + [x^3], 
\nonumber  \\
\fl \quad S_0\, \ln(x)^4 \, + [x]\, \ln(x)^3 \, + [x^2]\, \ln(x)^2 \, + [x^2]\, \ln(x) \, + [x^3],
 \nonumber  \\
\fl \quad S_0\, \ln(x)^3 \, + [x^2]\, \ln(x)^2 \, + [x^2]\, \ln(x)\,  + [x^3],
 \nonumber  \\
\fl \quad S_0\, \ln(x)^2 \, + [x^2]\, \ln(x) \, + [x^3], 
\nonumber  \\
\fl \quad S_0\, \ln(x) \,  + [x^3], 
\nonumber  \\
\fl \quad  S_0 \, \,=\,\, \,  \, \, 
1\,\, \,  +{\frac {1}{84}}\,{x}^{2}\,\, \,  +{\frac {5}{1764}}\,{x}^{3}\, \, \, 
+{\frac {263}{197568}}\,{x}^{4}\,\, \,  +{\frac {1355}{2074464}}\,{x}^{5}
\, \, \,\,  + \,\,  \cdots 
\end{eqnarray}
where the notation $\, [x^k]$ indicates an analytical series that begins as $\,\, x^k \, + \cdots$

\vskip 0.1cm
There are three (analytical at $0$) solutions.
The series $\, S_0$ corresponds to $\, LGF_7(x)$ and has the ``integrality 
property''~\cite{2013-rationality-integrality-ising}, i.e. it becomes, with the rescaling
$\, x \, \rightarrow \, 84 \,x$, a series with {\em integer coefficients}
\footnote[1]{This ``integrality property'' is expected, the series $\, LGF_7(x)$ being 
the diagonal of a rational function, it is 
{\em globally bounded}~\cite{2013-rationality-integrality-ising,Christol}.}.
With the rescaling $84$ which is the number of nearest neighbors in the lattice, the 
generating function counts the excursions in the fcc lattice. 
In contrast the series $\, S_2$ and $\, S_1$ are not 
{\em globally bounded}~\cite{2013-rationality-integrality-ising, Christol}: 
they cannot become series with integer coefficients up to a rescaling.

The occurrence of more than one non-logarithmic solution, at any regular singularity, may be
an indication that the linear differential operator factorizes. This is neither necessary
nor sufficient.

Since $\, G_{11}^{7Dfcc}$ is obtained with the series (\ref{seriesLGF7}) which is the series
in front of the log with the maximum power, 
if $\, G_{11}^{7Dfcc}$ factorizes, it should have a left factor of order seven.

\vskip 0.1cm

To factorize large order linear differential operators, we make use of the method sketched in 
Section 4 in~\cite{2009-chi5} (see also Section 6 in~\cite{{2014-SpecGeom-Ising}}).
This amounts to ``following'' the series pertinent to a specific local exponent
at a given singular point. 
For instance, we translate $\, G_{11}^{7Dfcc}$ to the point $\, x=\, -7$,
 and seek for the linear ODE corresponding 
to the series that begins as $\, t^2\,  + \cdots$ or as $\, t^{5/2}\,  + \cdots$, 
where $\,t= \, x\, +7$.
Linear combinations of series with different local exponents are considered as well, like
 the series $\, t^{3/2} \, + \cdots$ and the series 
$\, t^{5/2}\,  + \cdots$ at the point $\, x=\, -7$. 

Here we take the expansion around $\,x=\,0$, where we have the combination of three solutions
to consider.
In this method, we produce the general (analytic at 0)
solution of $G_{11}^{7Dfcc}$ which begins as
\begin{eqnarray}
\label{formalG11}
\hspace{-0.95in}  \quad  \quad  \quad 
 a_0 \,\,\,  + a_1\,x\,\, \, + a_2\,{x}^{2}\, \,\, + \, \Bigl({\frac {8401}{1746360}}\, a_0\, \, 
+{\frac {5581}{127008}}\, a_1\,\,  -{\frac {493}{2970}}\, a_2\Bigr)\cdot \,  {x}^{3}
\nonumber \\
\hspace{-0.95in}  \quad  \qquad 
+\, \Bigl({\frac {208069}{20321280}}\, a_1\,\, 
 -{\frac {70601}{488980800}}\, a_0\, \, 
+{\frac {360763}{2910600}}\, a_2\Bigr)\cdot \,  {x}^{4} 
\\
\hspace{-0.95in}  \quad  \qquad 
+\, \Bigl({\frac {903332869}{128024064000}}\, a_1\, -{\frac {146728781}{13335840000}}\, a_2\, 
+{\frac {878428781}{1120210560000}}\, a_0\Bigr)\,\cdot \,   {x}^{5} \,\, \, \, 
+\, \cdots
\nonumber
\end{eqnarray}
The expansion of $\, LGF_7(x)$, given in (\ref{seriesLGF7}), corresponds 
to the values $\, a_0=\, 1$, $\, a_1=\, 0$, $\, a_2=\, 1/84$.
The series $\, S_2$ corresponds to 
$a_0=\, 0$, $\, a_1=\, 0$, $\, a_2=\, 1$, and the series
$\, S_1$ corresponds to $\, a_0=\, 0$, $\, a_1=\, 1$, $\, a_2=\,-55/1008$.
The computation being done {\em modulo a prime}, the coefficients of the combination are
in the {\em finite range} $\,[1, p_r]$, $\, p_r$ being the prime.

The ``optimal'' linear ODE corresponds to the triplet $\, (Q_0=16, D_0=22, f_0=3)$.
For a given combination of the $a_j$, the series (\ref{formalG11}) is used to obtain
the ODE. If the parameter $f$ is found such that  $\, f > \, f_0=3$, 
this means that there is a right factor to $\, G_{11}^{7Dfcc}$.
If, for any combination of solutions (i.e. the coefficients $\, a_0$, $\, a_1$ and $\, a_2$ 
have been fixed to any value in the range $\,[1, p_r]$), one obtains $\, f=\, f_0$ 
there is no right factor over the rationals.
All the combinations have given the ``ODE-formula'' (\ref{ODEformLGF7}).

We have also shifted $\, G_{11}^{7Dfcc}$ to the point 
at infinity $ \,\,  x= 1/t= \, \infty$. Producing the series 
that begins as $\,\, t^{7/2} \cdot \,  ( 1\,  + \cdots )$
we find that the linear ODE is of {\em minimal order eleven}. Similarly, 
at the point $\, x=\, -7$, we find that 
the series beginning as (with $\, t=\, x\, +7$), 
$\,\, t^2 \cdot\,  (1 \, + \cdots )$, which corresponds to the local exponent 
$\, \rho=\, 2^3$, as well as the combinations of $\,\, t^{3/2} \cdot \,  (1\,  +\cdots )$ 
and $\,\, t^{5/2} \cdot \, (1 \, +\cdots )$, all require the linear differential 
operator $\, G_{11}^{7Dfcc}$. For all the other regular singularities, the 
non analytical local exponent is $\rho=\, 5/2$. This 
includes the point $\, x=\, 1$, for which the linear
ODE is shifted, and the corresponding series requires a linear ODE of order eleven.
We conclude that the order-eleven linear differential operator $\, G_{11}^{7Dfcc}$ 
is {\em irreducible} over the rationals.
Similar calculations show that $\, G_{6}^{5Dfcc}$ and $\, G_{8}^{6Dfcc}$ are also 
irreducible. For these two cases, we have carried out the check only 
with the analytical solutions around the point $\, x=\, 0$.

\vskip 0.1cm

Besides the irreducibility of all the known linear ODEs of the fcc lattice Green functions, 
one remarks that their singularities ``proliferate'' on the real axis as $\, d$ 
increases, and these singularities are {\em all rational numbers}. 
All the series $\, LGF_d(x)$, ($d \le\,  7$) have unity as convergence radius, 
and all the singularities of the linear ODE
are, besides $\, x = \, 1$, {\em outside} the unit circle $ \, \vert x \, \vert =\, 1$.
We may wonder whether these properties continue for the next $\, d$'s ?
In section \ref{landau}, we will show that it is not necessary to actually 
obtain the linear ODE to get the answer.
In the following, we show another property shared by 
the known linear ODE of the fcc lattice Green functions. 

\section{Canonical decomposition of the irreducible operator $\, G_{11}^{7Dfcc}$ }
\label{decomp}

In a previous paper~\cite{2014-DiffAlg-LGFCY},  
we have shown, for many operators, the equivalence of two properties: 
the {\em homomorphism of the operator with its adjoint}, and the occurrence of a 
{\em rational solution} for the {\em symmetric (or exterior)} square, or a drop of 
order of the symmetric (or exterior) square. The 
operators with these properties are such that their differential Galois 
groups are included in the symplectic or orthogonal differential groups. 
The differential Galois groups of the differential 
operators $\, G_{6}^{5Dfcc}$ and $\, G_{8}^{6Dfcc}$ are,
included respectively, in the $\, SO(6, \, \mathbb{C})$ and $\, Sp(8, \, \mathbb{C})$
groups.

In a recent paper~\cite{2014-Decomp-Special} we forwarded a ``canonical decomposition''
for those operators whose {\em differential Galois groups}
 are included in {\em symplectic or orthogonal}  groups.
These linear differential operators are homomorphic to their adjoints, and a ``canonical decomposition''
of these operators can be written in terms of a ``tower of intertwiners''~\cite{2014-Decomp-Special}.

For the differential operator $\, G_{6}^{5Dfcc}$, this canonical
decomposition reads (see eq.(41) in \cite{2014-DiffAlg-LGFCY}),
\begin{eqnarray}
\label{decomG6}
\hspace{-0.95in}&& \quad   \quad  \quad \quad \quad
G_{6}^{5Dfcc} \, \,  = \, \, \,\,   (V_{2} \cdot \,V_{1}\,\, + 1 ) \cdot r_1(x),   
\end{eqnarray}
where $\, r_1(x)$ is a rational function, and where 
$\,V_2$ and $\,V_1$ are {\em self-adjoint} operators of order, respectively, one and five.

The canonical decomposition of the operator $\, G_{8}^{6Dfcc}$ 
is (see eq.(61) in~\cite{2014-DiffAlg-LGFCY})
\begin{eqnarray}
\label{decomG8}
\hspace{-0.95in}&& \quad   \quad \quad  \quad \quad
G_{8}^{6Dfcc} \, \,  = \, \, \,\,   (W_{2} \cdot \,W_{1}\,\, + 1 ) \cdot r_2(x),   
\end{eqnarray}
where $\, r_2(x)$ is a rational function, 
and where $\,W_2$ and $\,W_1$ are {\em self-adjoint} operators of order, respectively, two and six.

In the sequel, we address this issue for the order-eleven linear differential 
operator $\, G_{11}^{7Dfcc}$. Note that using the formal solutions of the linear 
differential operator $\, G_{11}^{7Dfcc}$, it is easy to show that its 
{\em symmetric square} is of order $\,  65$ instead of the generically expected order $\, 66$. 
The operator $\, G_{11}^{7Dfcc}$ is in the differential Galois group $\,SO(11, \, \mathbb{C})$.

We find that the order-eleven linear differential operator $\, G_{11}^{7Dfcc}$ has 
the following decomposition:
\begin{eqnarray}
\label{canon-decompL11}
\hspace{-0.95in}&& \quad   \quad   
G_{11}^{7Dfcc} \, \,  = \, \, \,\,
  (U_{5} \cdot \,U_{4} \cdot \, U_{3} \cdot \, U_{2} \cdot \,  U_{1}
\,\,\, + U_{5} \cdot \, U_{4} \cdot \, U_{1} \,  \,
+ \, U_{5} \cdot \, U_{2} \cdot \,  U_{1} \, 
\nonumber \\
\label{last3}
\hspace{-0.95in}&&  \qquad \quad   \quad \, \, \,  \quad \quad    \quad 
 + \, U_{5} \cdot \, U_{4} \cdot \,  U_{3}
+ \, U_{3} \cdot \, U_{2} \cdot \, U_{1} 
\,\,\,\,  + U_{1} \, \, + \, U_{3}\,+ \, U_{5}) \cdot r(x),  
\end{eqnarray}
where $\, r(x)$ 
is a rational function, where $\, U_2$, $\, U_3$, $\, U_4$ and $\, U_5$ 
are {\em order-one self-adjoint} operators, and where $\, U_1$
is an {\em order-seven self-adjoint} operator. The $\, U_n$'s are 
too large to be given here.

The formula (\ref{canon-decompL11}) occurs because $\, G_{11}^{7Dfcc}$ is
{\em non-trivially homomorphic to its adjoint}
 \begin{eqnarray}
\hspace{-0.95in}&& \quad   \quad \quad     \quad  \quad \quad 
adjoint(L_{10}) \cdot \, G_{11}^{7Dfcc} 
\, \,  \, = \, \, \,  \,   adjoint\left(G_{11}^{7Dfcc}\right) \cdot \, L_{10},  
\end{eqnarray}
and the decomposition is obtained through a sequence of Euclidean rightdivisions.
Denoting $\, G_{11}^{7Dfcc}\,  = \, \, \, L_{[5]}$ and 
$\, L_{10}\,  = \, \, \, L_{[4]}$, one has
\begin{eqnarray}
\label{euclid}
\hspace{-0.95in}&& \, \,  \,    
L_{[5]} \,  = \, \, \, U_{5} \cdot \,L_{[4]}\, + \, L_{[3]}, \quad \, \,  
L_{[4]} \,  = \, \, \, U_{4} \cdot \,L_{[3]}\, + \, L_{[2]},  \quad  \, \, 
L_{[3]} \,  = \, \, \, U_{3} \cdot \,L_{[2]}\, + \, L_{[1]},  
\nonumber  \\
\hspace{-0.95in}&& \,\,  \,  \qquad   
L_{[2]} \,  = \, \, \, U_{2} \cdot \,L_{[1]}\, + \, L_{[0]}, \quad \, \quad 
L_{[1]} \, = \, \, U_{1} \cdot \, r(x), \quad \, \quad  
L_{[0]} \, = \, \, r(x).
\end{eqnarray}

A simple substitution gives the decomposition (\ref{canon-decompL11}) in terms
of the five self-adjoint operators $\, U_n$.

From the decomposition (\ref{canon-decompL11}), and since the self-adjoint operator $\, U_{1}$ is 
of {\em odd order  greater than one} (see~\cite{2014-Decomp-Special} for the details), 
one can understand why the {\em symmetric square} of the order-eleven linear differential operator 
$\,G_{11}^{7Dfcc}$ is of order $\, 65$ instead of the generically expected order $\, 66$.  
The symmetric square of the self-adjoint order-seven operator $\, U_{1}$ is of order 27
instead of the generically expected order 28.

From the decomposition (\ref{canon-decompL11}) one can also deduce that the 
{\em symmetric square of the adjoint} of the order-eleven linear 
differential operator $\, G_{11}^{7Dfcc}$ 
has a {\em rational solution} that is the square of the solution 
of the adjoint of the order-one operator $\, U_5$. 

One can also see the second intertwining relation between the order-eleven linear differential 
operator $\, G_{11}^{7Dfcc}$ and its adjoint
\begin{eqnarray}
\hspace{-0.95in}&& \quad     \quad  \quad   \quad \quad \quad 
 G_{11}^{7Dfcc} \cdot \, L_4 \, \,\, = \, \, \, \,
 adjoint(L_4)  \cdot \, adjoint\left(G_{11}^{7Dfcc}\right).  
\end{eqnarray}

Again, one can consider the homomorphisms of $\, L_4$ with its adjoint,
and  the homomorphisms of  $\, L_{10}$ with its adjoint, getting this way a 
``tower of intertwiners'' (see~\cite{2014-Decomp-Special}).
In terms of the $\, U_n$'s, they read 
\begin{eqnarray}
\label{L4decom}
\hspace{-0.95in}&& \quad  \quad   
 L_4 \, \,\,  = \, \, \, 
{{1} \over {r(x)}} \cdot \, (U_{2} \cdot \, U_{3} \cdot \, U_{4} \cdot \,  U_{5}
\, + U_{2} \cdot \, U_{5} \, 
+ \, U_{2} \cdot \,  U_{3} \, 
+ \, U_{4} \cdot \,  U_{5} \, + \, \, 1),  
\end{eqnarray}
and
\begin{eqnarray}
\label{L10decom}
\hspace{-0.95in}&& \quad    \quad 
 L_{10} \, \,\,  = \, \, \, \,  (U_{4} \cdot \, U_{3} \cdot \, U_{2} \cdot \,  U_{1}
\, + U_{4} \cdot \, U_{1} \, 
+ \, U_{4} \cdot \,  U_{3} \, 
+ \, U_{2} \cdot \,  U_{1} \, + \, \, 1) \cdot \,  r(x). 
\end{eqnarray}
These two intertwiners have ``special'' differential Galois groups, included in 
$\, SO(4, \, \mathbb{C})$ and $\, SO(10, \, \mathbb{C})$ for respectively 
$\, L_4$ and  $\, L_{10}$.

The symmetric square of $\, L_4$ has a rational solution corresponding to the square of the
solution of $\, U_5$. The symmetric square of the adjoint of $\, L_4$ annihilates a rational
solution which is the square of the solution of the adjoint of $\, U_2$.

The symmetric square of $\,L_{10}$ has a drop of order: it is of order $\, 44$,
instead of the order $\, 45$, generically expected for an order-ten operator.
The symmetric square of the adjoint of $\, L_{10}$ is of the generic order $\, 45$
and annihilates a rational solution which is the square of the solution of the adjoint
of $\, U_4$.

\vskip 0.1cm

\section{Landau singularities}
\label{landau}

There is an analytical approach to obtain the singularities of 
functions, defined through $\, n$-fold integrals, which amounts to
 imposing {\em conditions on the integrand}. 
These conditions on the integrand are called 
{\em Landau  
conditions}~\cite{2007-PhiD-integrals,2007-PhiH-integrals,landau-1959,bjor-drel-65,ed-la-ol-po-66,itz-zub-80,nickel-99,
nickel-00,nickel-05}, and the singularities obtained this way are called 
{\em Landau singularities}.

The singularities of the linear ODE, annihilating the $ \, d$-dimensional 
face-centred cubic lattice Green functions, 
are obtained by solving, in the variables $\, \cos(k_j)$, the {\em Landau 
equations}\footnote[3]{See section 2 in~\cite{2007-PhiD-integrals} for an example of Landau equations
for a simple integrand as in (\ref{LGFd}).} corresponding to (\ref{LGFd})
\begin{eqnarray}
\label{eqnLandau}
\hspace{-0.95in}&& \qquad \qquad 
(1-\cos(k_j)^2) \cdot \, 
\sum_{i \ne j}^d \cos(k_i) \,\,=\,\,\, 0, 
 \qquad  \quad  j=\, 1,\,  2,\, \cdots,\,  d,
\end{eqnarray}
and plugging in: 
\begin{eqnarray}
\label{thex}
\hspace{-0.95in}&& \qquad \qquad \qquad 
 x_s \, \, = \, \, \,
 { d \choose 2} /
 \Bigl(\sum_{i=1}^d \, \sum_{j=i+1}^d \cos(k_i) \cdot \, \cos(k_j)\Bigr).
\end{eqnarray}

It is straightforward to solve the system (\ref{eqnLandau}) for any value of $\, d$.
When all $\, \cos(k_j)^2 = \, 1$, the system of equations (\ref{eqnLandau}) is verified, and
one obtains the singularity $\, x_s=\, 1$. For $\, \cos(k_1)^2=\, 1$
 and $\, \cos(k_j)^2 \ne\,  1,\,  j \ne\,  1$, the
system (\ref{eqnLandau}) becomes linear, and one obtains a set of singularities. The next step is
to consider $\cos(k_1)^2=\, 1$, $\cos(k_2)^2=\, 1$ and $\, \cos(k_j)^2 \ne\,  1,\,  j >\,  3$, a set
of singularities (that may include some of the first set) is obtained, and so on.
At each step, the problem amounts to solving a linear system where the coefficients are $\, 0$ and $\, 1$.

\vskip 0.1cm

The set $\, x_s$ of singularities read
\begin{eqnarray}
\label{thexs}
\hspace{-0.95in}&& \qquad \qquad  \qquad  \quad 
  x_s \,\,  =\,\, \, \,  { d \choose 2} \cdot \, {\frac{1}{ \xi(d, k, j)}}, 
\end{eqnarray}
where
\begin{eqnarray}
\label{singfirstset}
\fl \qquad 
\xi(d, k, j)\,  \,=\, \, \, \, \,
{\frac{{d}^{2}\, \, 
- \, (k \, +4\,j\, +1)  \cdot\,  d\,\, +4\,{j}^{2}\,+k\,+4\,j\,k }{2\, \,(1-k) }}, \\
\fl \qquad 
 \quad \quad   {\rm with} \quad  \quad  \quad 
k=\, 0,\,  2, \, 3,\,  \cdots,\,  d-1, \quad  \quad  \quad  j=\, 0, \, \cdots,\,  [(d-k)/2],  
\nonumber
\end{eqnarray}
and where $\, [x]$ is the integer part of $\, x$. 

\vskip 0.1cm

For $\, d=\, 4$,  we obtain the regular singularities 
$\, x_s =\,  -8,\,  -6, \, -3, \, -2, \, 1$, which, indeed, occur
in the linear differential equation obtained by Guttmann~\cite{guttmann-2010}.
For $\, d=\, 5$, the regular singularities are 
$\, x_s =\,  -15,\,  -10,\,  -5, \, -5/3,\,  1,\,   5$, in agreement with
the  linear differential equation obtained by Broadhurst~\cite{broadhurst-2009}.
One also has agreement with the singularities of the  linear
differential equation obtained by  Koutschan~\cite{koutschan-2013}
 for $\, d=\, 6$: our Landau singularities 
read 
$\, \, x_s =\,  -24,\,  -15,\,  -9,\,  -60/7,$
$\, -15/2, \, -5,\,  -4,\,  -15/4, \, -3/2,\,  1, \, 3$.

\vskip 0.1cm

For $\, d=\, 7$, our Landau singularities (\ref{thexs}) 
are in agreement with the singularities occurring in the linear ODE and given in (\ref{singLGF7}). 

\vskip 0.1cm

The Landau singularities corresponding to $\, LGF_d(x)$, and given by (\ref{thexs}), are, 
obviously, {\em all rational numbers}, and are all 
such that $\, \vert x_s \vert  > 1$, except of the
singularity $\, x_s=\, 1$. 

\vskip 0.1cm

As the dimension $\, d$ goes higher, the number of the singularities goes as $\, d^2$.
For each dimension $\, d$, the minimal value of the singularities 
is given by $\, S_{min}=\, -d \cdot \, (d-2)$,
while the maximum value $\, S_{max}$ of the singularities depends on whether 
the dimension $\, d \, $ is (or how far is it from) a perfect square. For instance, 
\begin{eqnarray}
\label{theSmax}
\hspace{-0.95in}&& \quad \quad 
 \sqrt{d} \, = \, {\rm integer}, \qquad \quad \quad \,\, \,\,   S_{max}\,=\, 
{1 \over 4}\, d \cdot \, (\sqrt {d}- 1), \\
\hspace{-0.95in}&& \quad \quad  \sqrt{d+1} \,= \, {\rm integer}, \qquad \quad 
S_{max}\,=\, \,\, {1 \over 2}\,  (d-1) \cdot \, (\sqrt {d+1} \, -1),  \\
\hspace{-0.95in}&& \quad \quad   \sqrt{d-1} \, = \, {\rm integer}, \qquad \quad 
S_{max}\,=\,\,\,  {1 \over 2}\, d  \cdot \, \sqrt {d-1}, \\
\hspace{-0.95in}&& \quad \quad   \sqrt{d+2} \, = \, {\rm integer}, \qquad \quad 
S_{max}\,=\, \,\, {1 \over 2}\, d \cdot \, (d-1), \\
\hspace{-0.95in}&& \quad \quad    \sqrt{d-2} \, = \, {\rm integer}, \qquad \quad 
S_{max}\,=\, \, \,
{1 \over 2}\,{\frac { d \cdot \, (d-1)  \cdot \, (2\,\sqrt {d-2}\, -1) }{4\,d \, -9}}.
\end{eqnarray}
To show the deviation from the perfect square, we define the parameter $d$ as
\begin{eqnarray}
d\,  \,=\,\,  n^2 \,+\, p, \qquad \qquad  0 \le\, \,  p \,  \le \, \,  2\,n.
\end{eqnarray}
From inspection on many $S_{max}$, we infer
\begin{eqnarray}
\label{Smax}
S_{max} \, \,=\, \,\, \, \, 
 {\frac{(n^2+p)\cdot \, (n^2\, +p\, -1)}{
3\, n \, - p \,\,  + {5 \over 2}\, \,  +(-1)^p \cdot \, (n+ {3 \over 2})}}, 
\end{eqnarray}
which corresponds to the singularity (\ref{thexs}), with (\ref{singfirstset}) evaluated for $k =0$
and the following value of $j$
\begin{eqnarray}
\label{jdn}
 j \,\,  = \, \, \,\, \,   
{d \over 2} \,\, \,  - {n \over 2}\, \, \,  - {3 \over 4} \, \,\,  - {(-)^p \over 4}.
\end{eqnarray}

\section{From facts to speculations}
\label{specul}

\subsection{The order of the linear ODE of the $d$-dimensional fcc lattice}
\label{order}

Let us consider the local exponents at the three singularities $\, x=\, 0,\,  \infty,\,  1$,
common to all the linear ODE for $\, d=\, 3,\,  \cdots,\,  7$.
One remarks\footnote[1]{The notation (e.g.) $\, 2^2$ means that $\, \rho=\, 2$ occurs twice.}
(see Table \ref{Ta:1}) the simple pattern at $\, x\, =\infty$ and $\, x=\, 1$, 
which depends on the {\em parity} of $\, d$. This leads to imagine, for $\, d=\, 8$, 
that one may have, respectively, $\, \rho=\, 4^2$ and $\, \rho=\, 3^2$.
More important is the pattern at the regular singularity $\, x=\, 0$, which, if well guessed,
should give $\, \rho = \, 0^8,\,  1^4, \, 2^2$, for the next $\, d=\, 8$. This 
would give an order-14 linear ODE for $d=8$.
For $\, d=\, 9$, the pattern may give $\, \rho =\,  0^9,\,  1^5,\,  2^3, \, 3$, 
indicating an order-18 linear ODE, and for $\, d=\, 10$, it may be 
$\, \rho = \, 0^{10},\,  1^6,\,  2^4,\,  3^2$, i.e. an order-22 linear ODE, etc.

\begin{table}[htdp]
\caption{
The local exponents $\, \rho$ at $\, x=\, 0$, $\, x=\, \infty$ and $\, x=\, 1$.
For the last two points, only the non analytical exponent is given.
 }
\label{Ta:1}
\begin{center}
\begin{tabular}{|c|c|c|c|}\hline
    $d$ &  $x=0$ &  $x=\infty$ & $x=1$        \\ \hline 
\hline
    3 & $0^3$          &  $3/2$    &  $1/2$    \\
    4 & $0^4$          &  $2^2$    &  $1^2$     \\
    5 & $0^5, 1$       &  $5/2$    &  $3/2$     \\
    6 & $0^6, 1^2$     &  $3^2$    &  $2^2$    \\
    7 & $0^7, 1^3, 2$  &  $7/2$    &  $5/2$     \\
\hline
 \end{tabular}
\end{center}
\end{table}
If all this is correct, the order of the ODE of the $\, d$-dimensional fcc lattice 
should read:
\begin{eqnarray}
\label{orderq}
q \, \,=\,\,\, \,  {\frac{d^2}{4}} \,\,\, 
   - {\frac{d}{2}}\, \, \, + {\frac{17}{8}} \,\,\,  - {\frac{(-)^d}{8}}. 
\end{eqnarray}

\begin{table}[htdp]
\caption{
 The number of terms $\, N_m$ (and $\, N_0$) needed 
to obtain the minimal order ODE (and the optimal order ODE)
annihilating $\, LGF_d(x)$.
The numbers in the row for $\, d=\, 8$ are estimates.
 }
\label{Ta:2}
\begin{center}
\begin{tabular}{|c|c|c|c|c|}\hline
    $d$ &  $N_m$ &  $N_0$ & $N_m-N_0$ & $Q_{opt}-Q_{min}$       \\ \hline 
\hline
    4 & 40  & 40   & 0 & 4-4=0     \\
    5 & 98  & 88   & 10 & 7-6=1     \\
    6 & 342  &  228  & 114 & 11-8=3    \\
    7 & 732  &  391  & 341 & 16-11=5     \\
    8 & (1650)  & (672)   & (978) &(20-14=6)     \\
\hline
 \end{tabular}
\end{center}
\end{table}

For the linear ODE of the eight-dimensional fcc lattice, if we assume the order to be 14,
the number of the singularities given by (\ref{thex}) as 19 (omitting the multiplicity),
and the degree of the apparent polynomial twice the degree of $\, P_{11}$, one obtains 
the estimates given in Table \ref{Ta:2}.

 \vskip .1cm

\subsection{The differential Galois group of the linear ODE for $\, LFG_d$}
\label{observ}

We have seen that the differential Galois groups of the linear differential operators
for the fcc lattice Green functions in five, six and seven dimensions, namely 
$\, G_{6}^{5Dfcc}$, $\,G_{8}^{6Dfcc}$ and $\,G_{11}^{7Dfcc}$ are respectively included
in the $\,SO(6, \, \mathbb{C})$, $\,Sp(8, \, \mathbb{C})$ and 
$\,SO(11, \, \mathbb{C})$ group. From the previous speculations on the minimal-order
ODEs for the $\, d$-dimensional fcc lattice Green functions, one could imagine that 
the differential Galois groups of the linear differential operators
for the fcc lattice Green functions in eight and nine dimensions, namely 
$\, G_{14}^{8Dfcc}$ and $\, G_{18}^{9Dfcc}$ would be included respectively in 
$\,Sp(14, \, \mathbb{C})$ and $\,SO(18, \, \mathbb{C})$.

It has been underlined in~\cite{2014-Decomp-Special} that 
the right-most self-adjoint operator 
in the ``canonical decomposition'' of these operators with ``special'' differential 
Galois groups plays a selected role: the rational solutions of the symmetric 
(resp. exterior) square of these operators (or the drop of order of these squares)
{\em does not} depend on the other self-adjoint operators of the decomposition but {\em only} 
on this right-most self-adjoint operator (multiplied by the rational function $\, r(x)$
of the decomposition). This right-most is $\, V_1$ for $\,G_{6}^{5Dfcc}$ (see (\ref{decomG6})),
$\,  W_1$ for  $\,G_{8}^{6Dfcc}$ (see (\ref{decomG8})), and $\, U_{5}$ 
for  $\,G_{11}^{7Dfcc}$ (see (\ref{canon-decompL11})). Let us remark that 
the right-most self-adjoint operators of the linear differential operators
for the fcc lattice Green functions in five, six and seven, are respectively of order
five, six and seven. It is, thus, tempting to imagine that the linear differential operators
for the fcc lattice Green functions in eight and nine dimensions, will correspond to a
decomposition generalizing (\ref{canon-decompL11}) (see~\cite{2014-Decomp-Special}), 
involving three order-two self-adjoint operators and a right-most self-adjoint operator
of order eight, and that the decomposition for the operator associated with the 
fcc lattice Green function in  nine dimension will correspond to nine 
order-one self-adjoint operators and a right-most self-adjoint operator
of order nine. A similar ``ansatz'' can be made for fcc lattice Green 
functions in any dimension $\, d$: we could imagine decompositions in terms of 
a right-most self-adjoint operator of order $\, d$
and $\, q \, -d$ 
order-one self-adjoint operators for $\, d$ odd, and 
$\, (q \, -d)/2$ order-two self-adjoint operators  for $\, d$ even,
where $\, q$ is given by (\ref{orderq}).

\subsection{Return probability}

The latttice Green functions arise in the study of random walks in probability theory.
Here the value at $\, x=\, 1$ of $\, LFG_d(x) =\,  P_d(1)$ gives the {\em return probability}
\begin{eqnarray}
\label{Rd}
\hspace{-0.75in}&& \quad \quad \qquad
R_d  \,\, = \,\, \, \,  1 \,\,  - {\frac{1}{P_d(1)}}, 
\end{eqnarray}
which is the probability for a walker, starting at the origin, to return to the origin.

For the $\, d=\, 3 \, $ fcc lattice, the return probability is known 
in closed form~\cite{watson-1939}
\begin{eqnarray}
\label{R3}
\hspace{-0.75in}&& \quad \quad \qquad
R_3 \,\,  =\, \, \, \, 
1 \,\, \,  - \, {\frac{16 \,\,  4^{1/3} \,\,  \pi^4 }{9 \,\,  \Gamma(1/3)^6}},  
\end{eqnarray}
while, for $\, d=\, 4,\,  \cdots,\,  6$, only the numerical values are obtained.
With the recursion of the order-eleven ODE annihilating $\, LGF_7(x)$, one computes the 
number of terms of the series necessary to evaluate $\, R_7$ to the desired accuracy.
The value\footnote[2]{$R_7 = \, 0.017\,563\,245\,036\,917\,347\,782\,481\,698\,$
$942\,709\,061\,089\,911\,556\,997\,564\,570\,73 \, \cdots$}
 of $\, R_7$ completes the Table given in~\cite{koutschan-2013},
which shows a dependence on the dimension of the lattice.
The values of $\, R_d$, in Table \ref{Ta:3}, shows a decrease, as $\, d$ goes higher,
that can, for instance, be of the form $\, d^{-\alpha}$, with $\alpha > 0$.

\begin{table}[htdp]
\caption{
 The return probability as a function of the dimension.
 }
\label{Ta:3}
\begin{center}
\begin{tabular}{|c|c|}\hline
    $d$ &  $R_d$ (10 digits)       \\ \hline 
\hline
    2 & 1      \\
    3 & 0.256 318 236 5     \\
    4 & 0.095 713 154 1      \\
    5 & 0.046 576 957 4      \\
    6 & 0.026 999 878 2     \\
    7 & 0.017 563 245 0      \\
\hline
 \end{tabular}
\end{center}
\end{table}

\vskip 0.1cm

The recursive method of generation of series of section \ref{generationseries} 
allows us to produce the first 60 terms of the series up to the dimension $\, d=\, 40$, 
from which we infer the expansion of (\ref{LGFd}) for the {\em generic} dimension 
$\, d$ up to $\, x^{36}$.
The first eight terms of $\, LGF_d(x)$ read:
\begin{eqnarray}
\label{seriesLGFd}
\hspace{-0.95in}&& \quad \quad  
   LGF_d(x) \, \,= \,\, \,\,  1\,\,\,\,  +{\frac {1}{2 \, \, (d-1) \cdot \, d}} \, x^2\, \,\, \, 
+{\frac { \left( d-2 \right) }{ (d-1)^{2} \cdot \, {d}^{2}}}\cdot  \, x^3\, \,
   \\
\hspace{-0.95in}&& \quad \quad \quad \quad \quad 
 +\, {\frac { 3 \cdot \, (10\,{d}^{2}-38\,d+39) }{8 \,  \,  {d}^{3} \, (d-1)^{3}}}\cdot  \, {x}^{4}
\, \,\,  
+  {\frac { (d-2)  \cdot \, (34\,{d}^{2}-148\,d+183) }{2  \, \,  {d}^{4} 
 \cdot\, (d-1)^{4}}} \cdot \, {x}^{5}
\nonumber \\
\hspace{-0.95in}&& \quad \quad \quad  \quad \quad   
+{\frac {5 \cdot \, (302\,{d}^{4}\, -2824\,{d}^{3}\, +10357\,{d}^{2}\, -17417\,d\, +11176) }{
16 \, \,  {d}^{5} \cdot \, (d-1)^{5}}} \cdot \,{x}^{6}  
\nonumber\\
\hspace{-0.95in}&& \quad \quad  \, \, 
+{\frac {3 \,\,  (d-2)  \,
(1646\,{d}^{4}\, -17868\,{d}^{3}\, +77749\,{d}^{2}\, -158367\,d\, +125870) }{
8  \,  {d}^{6} \cdot  \, (d-1)^{6}}} \cdot \,{x}^{7}
 \,\,\, \,  + \,\cdots, 
 \nonumber
\end{eqnarray}
and has the form
\begin{eqnarray}
   LGF_d(x) \, \,= \,\, \, \, 
\sum_n \,\, {\frac{C_{n-2} (d)}{ (d \cdot \, (d-1))^{n-1}}} \cdot \, x^n, 
\end{eqnarray}
where the $\, C_n(d)$'s are {\em polynomials} in $\,d$ of degree $\,n$.

For large values of $ \,d$ one can deduce from $\, LGF_d(x)$, 
a series expansion in $ \, 1/d$, namely 
\begin{eqnarray}
\label{seriesLGFdbig}
\hspace{-0.95in}&& \quad \, 
   LGF_\infty(x) \, \,= \,\, \, \,  1\,\,\,
 + {{x^2} \over {2}} \cdot \, {{1} \over {d^2}} \, \,
+ \, \, {{x^2} \over {2}} \cdot  \, (2\, x \, +1)  \cdot  \, {{1} \over {d^3}}
\,\,\,  + \, \, {{x^2} \over {4}} \cdot  \, (15 \, x^2 \, +2)  \cdot  \, {{1} \over {d^4}}  
\nonumber \\ 
\hspace{-0.95in}&& \quad \quad  \quad  \quad \quad   \quad \,\, \, 
+ {{x^2} \over {2}} \cdot \, (34 \, x^3 \, -6 \, x^2 \, -2 \, x \, +1)  \cdot \, {{1} \over {d^5}}
 \, \, \, \, + \, \, \cdots 
\end{eqnarray}
where the fact that coefficients of $\, 1/d^n$ in (\ref{seriesLGFdbig})
are {\em polynomials} is a straight consequence of the fact that 
the (rational functions of $\, d$) coefficients of $\, x^n$ in  (\ref{seriesLGFd}) 
behave as $\, 1/d^n$ for large $\, d$.

One deduces with (\ref{seriesLGFdbig}), the expansion,
for large values of $d$, of the return probability $\, R_d$: 
\begin{eqnarray}
\label{Returnatinf}
\hspace{-0.95in}&& \quad \quad \quad \quad
 R_d  \,\,  \, =\, \,\, \,  \, 
 {\frac{1}{2\, d^2}}\, \, \,   + {\frac{3}{2\, d^3}}\,\,   +{\frac{4}{d^4}} \, \, + {\frac{12}{d^5}}
 \, + {\frac{327}{8\, d^6}}\, \,  + {\frac{1219}{8\, d^7}} \,\,  \,  \,  + \, \, \cdots,  
\end{eqnarray}
which has the ``integrality property'', i.e. it becomes a series with 
{\em integer coefficients} with the rescaling $\, d \rightarrow \, d/2$.

The series (\ref{Returnatinf}) is a {\em divergent}\footnote[2]{As far as it is 
safe to make the conclusion with a series of only 34 terms.} series,
 and shows the trend in $\, d^{-2}$ for large values of $\, d$.

\vskip 0.1cm

{\bf Remark:} The fact that the coefficients of $\, x^n$ in  (\ref{seriesLGFd}) behave 
as $\, 1/d^n$ for large $\, d$, {\em does suggest} to introduce the rescaled variable 
$\, y \, = \, 2 \, x/d$, into the series (\ref{seriesLGFd}) to obtain a series
in $\, y$ with rational coefficients in $\, d$. In the 
$\, d \,\rightarrow \, \infty$ limit this series becomes a (divergent) 
series with {\em integer} coefficients:
\begin{eqnarray}
\label{yseries}
\hspace{-0.95in}&&  \quad 1 \, \, +2\, y^2\,+8\,y^3\,+60\,y^4\,
+544 \,y^5\, +6040\,y^6\, +79008 \,y^7\,+1190672\,y^8 \, \,+ \, \cdots 
\end{eqnarray}
which is actually solution of a second order linear 
{\em non-Fuchsian} ODE\footnote[1]{
The corresponding order-two operator has the unique factorization  
$(d/dy+(3\,y+2)/(y ( y+1 ))) \cdot (d/dy+(2\,y^{2}-1)/(2 y^{2} ( y+1)))$.}
with the two solutions  
\begin{eqnarray}
\label{solunonFuchs}\hspace{-0.95in}&& \quad \quad
(y+1)^{-1/2} \cdot \, y^{-1/2} \cdot \, e^{-1/2/y}, \qquad 
  _1F_1\Bigl(\Bigl[ {{1} \over {2 }} \Bigr],\Bigl[  {{3} \over {2 }}  \Bigr],
 \, {{ y+1} \over {2  \,  y}} \Bigr)   \cdot \, e^{-1/2/y}.
\nonumber  
\end{eqnarray}

\section{Conclusion}
\label{summary}

In this paper we presented a recursive method to generate the series coefficients 
of the lattice Green function of the $d-$dimensional fcc lattice.
The series for $\,d=7$ has been obtained to some length 
and thanks to the idea that many linear ODEs, of order larger than 
the minimal one, may be obtained with {\em fewer terms}, we have obtained the 
linear ODE that annihilates the lattice Green function 
of the seven-dimensional fcc lattice.

\vskip 0.1cm

The (minimal-order) linear differential equation for the 
seven-dimensional fcc lattice is of {\em order eleven} and is irreducible.
We show that the differential Galois group of the corresponding operator is included
in $\, SO(11, \, \mathbb{C})$. This order-eleven operator is non-trivially 
homomorphic to its adjoint, and we found a decomposition of this operator in terms of
four order-one self-adjoint operators and one order-seven self-adjoint operator.

\vskip 0.1cm

As far as the singularities are concerned, we show that the singularities of 
$\, G_{11}^{7Dfcc}$ follow the scheme observed for the linear ODEs of lower $\, d$.
With the Landau conditions approach, we showed that the singularities, for {\em any} $\, d$, 
are all {\em rational numbers}, which are, beyond $\, x\, = \, 1$,  outside the interval 
$\, [-1, \, 1]$.

\vskip 0.1cm

As the return probability in random walks in (fcc) lattice may be of interest,
we also evaluated the return probability $\, R_7$ 
of the seven-dimensional fcc lattice.
The recursive method of generation of series terms of $\, LGF_d(x)$ is sufficiently
effective to let us produce short series for many values of $\,d$. 
With these results, we inferred that the return probability 
$\, R_d$  behaves as $\, d^{-2}$ for large values of $\, d$.

\vskip 0.1cm

The linear differential operators whose differential Galois
groups are included in the symplectic or orthogonal groups have 
a ``canonical decomposition'' in terms of self-adjoint 
operators~\cite{2014-Decomp-Special}. 
For the known linear differential operators $\, G_{6}^{5Dfcc}$, $\, G_{8}^{6Dfcc}$ 
and $\, G_{11}^{7Dfcc}$,
some observations on the specific ``ansatz'' form  
of their ``canonical decompositions'' in terms of self-adjoint operators are sketched in 
section \ref{observ}. 
These speculations are a strong incentive to perform the 
calculations for the fcc lattice Green function, in the higher 
dimensions\footnote[1]{For $d=8$, the time needed to generate 50, 100 and 150 terms 
is respectively 4, 59 and 331 seconds.
To generate 672 terms (see Table \ref{Ta:2}) plus 100 terms for the check, the time can 
be estimated as at least 82 hours on a desk computer.}.

\ack

We thank J-A. Weil for help on the calculations of section \ref{decomp}. 
This work has been performed without any support of the ANR, the ERC or the MAE. 

\appendix

\section{The apparent polynomial of $\, G_{11}^{7Dfcc}$}
\label{appendixApp}
\begin{eqnarray}
\fl \qquad P_{11} \, \,= \,  \, \,
4033844291292160512572197716389040432464926556160000000
 \nonumber \\
\fl \qquad -76716614566403110639033419440712752789656370047488000000\,x
 \nonumber \\
\fl \qquad +1036105712621875785358061549706636065154061705507660800000\,{x}^{2}
\nonumber \\
\fl \qquad +16582552474758671672601408908172980846455437334039680000000\,{x}^{3}
 \nonumber \\
\fl \qquad +92528281358446684721054852304052021993648722856652790720000\,{x}^{4}
 \nonumber \\
\fl \qquad +63191344208515083154464238671792397457218673838108305673600\,{x}^{5}
 \nonumber \\
\fl \qquad -1028373259267414517632638438252917863677228096786073040753280\,{x}^{6}
 \nonumber \\
\fl \qquad -3875346880791921428013321015467132142892901930768457490486272\,{x}^{7}
  \nonumber \\
\fl \qquad -6920318399426800817887962016170406919476899714148547999153088\,{x}^{8}
  \nonumber \\
\fl \qquad -8154471373075656935436312555783320414633120129789847214836048\,{x}^{9}
   \nonumber \\
\fl \qquad -6245702881422338462232184489205592634805720476083959916954720\,{x}^{10}
  \nonumber \\
\fl \qquad -2376232682826675746826787210063813304210071338288106265892144\,{x}^{11}
   \nonumber \\
\fl \qquad +409323467347919017193444730250431709768180749972825987763968\,{x}^{12}
   \nonumber \\
\fl \qquad +1094294121359871591653112881632669129584109878287736283862304\,{x}^{13}
   \nonumber \\
\fl \qquad +752324466393615486137798566063739247640534078467948502617792\,{x}^{14}
   \nonumber \\
\fl \qquad +369596809799205895313185288244531423864737167506726868047424\,{x}^{15}
   \nonumber \\
\fl \qquad +173462105486017718670525285173642124785703042463376755656000\,{x}^{16}
   \nonumber \\
\fl \qquad +81585560140780999074640514392799621227191196083376547520384\,{x}^{17}
   \nonumber \\
\fl \qquad +30039869312248151750092120565576302197789247371702033184576\,{x}^{18}
   \nonumber \\
\fl \qquad +3352864878089336012143385345711247821224483161365521297536\,{x}^{19}
  \nonumber \\
\fl \qquad -5269157075175728093621430002781092748292821974628349580736\,{x}^{20}
  \nonumber \\
\fl \qquad -5003037996223007820302755774000779739244471692540950704576\,{x}^{21}
  \nonumber \\
\fl \qquad -2724443564627932817554684016950743167044585588150372246848\,{x}^{22}
 \nonumber \\
\fl \qquad -1099950431710329286006976673908885512918136416974142913472\,{x}^{23}
 \nonumber \\
\fl \qquad -354827676457302481549017891474813922001077928917049120576\,{x}^{24}
  \nonumber \\
\fl \qquad -94649380054954516698535249719599292251951455673840066208\,{x}^{25}
  \nonumber \\
\fl \qquad -21283773506777832888307450022897921617853237366966188416\,{x}^{26}
  \nonumber \\
\fl \qquad -4083520971052879322404793887394253941827038658791531616\,{x}^{27}
  \nonumber \\
\fl \qquad -673847982635093459687405322938556215952418650082540224\,{x}^{28}
  \nonumber \\
\fl \qquad -96232684263265628624960727707621390080193285430960000\,{x}^{29}
 \nonumber \\
\fl \qquad -11983439352062136145880401362426806595412251158557632\,{x}^{30}
  \nonumber \\
\fl \qquad -1321477574928217810528106035645553007130680877071168\,{x}^{31}
  \nonumber \\
\fl \qquad -133756126413404887811038857155026506627716809584192\,{x}^{32}
  \nonumber \\
\fl \qquad -13277030500839992506125996150534276486354892529664\,{x}^{33}
  \nonumber \\
\fl \qquad -1384266419715171016988616613973267638084553135424\,{x}^{34}
  \nonumber \\
\fl \qquad -151391595348867999321027885040943286501123115008\,{x}^{35}
  \nonumber \\
\fl \qquad -16057106216119177077113061732259948384006651712\,{x}^{36}
  \nonumber \\
\fl \qquad -1518287617204750975914622379591236350411307968\,{x}^{37}
  \nonumber \\
\fl \qquad -121212363333027580459494623106043808245269312\,{x}^{38}
  \nonumber \\
\fl \qquad -7904889970969434550214545179960157702136640\,{x}^{39}
  \nonumber \\
\fl \qquad -410180899572074649344019059013918950937600\,{x}^{40}
  \nonumber \\
\fl \qquad -16438739943524974573089152060006574677904\,{x}^{41}
  \nonumber \\
\fl \qquad -488014173372889647550579109740889580576\,{x}^{42}
 \nonumber \\
\fl \qquad -10039688420130121713386066606706194160\,{x}^{43}
 \nonumber \\
\fl \qquad -126636870801841392088882506641659200\,{x}^{44}
 \nonumber \\
\fl \qquad -727932992011727859393396784221600\,{x}^{45}.  
\nonumber 
\end{eqnarray}

\end{document}